\documentclass[11pt, a4paper]{article}
\usepackage{jheppub}
\usepackage{amsmath}
\usepackage{amssymb}
\usepackage{epsfig}
\usepackage{graphicx}
\usepackage{hyperref}
\usepackage{tikz}
\usetikzlibrary{decorations.pathmorphing}
\usepackage{pgfplots}
\pgfplotsset{width=6cm}
\usepackage{graphicx}

\newcommand{\T}{\mathfrak{t}}
\newcommand{\A}{\mathcal{A}}
\newcommand{\C}{\mathcal{C}}

\newcommand{\OO}{\mathcal{O}}
\newcommand{\M}{\mathcal{M}}

\newcommand{\HO}{H_{\text{CFT}}}

\begin{document}
	\title{Divergences in the rate of complexification}
	\author{Mudassir Moosa}%
	\affiliation{ Center for Theoretical Physics and Department of Physics\\
		University of California, Berkeley, CA 94720, USA 
	}%
	\affiliation{Lawrence Berkeley National Laboratory, Berkeley, CA 94720, USA}
	\emailAdd{mudassir.moosa@berkeley.edu}
	\abstract{
	It is conjectured that the average energy provides an upper bound on the rate at which the complexity of a holographic boundary state grows. In this paper, we perturb a holographic CFT by a relevant operator with a time-dependent coupling, and study the complexity of the time-dependent state using the \textit{complexity equals action} and the \textit{complexity equals volume} conjectures. We find that the rate of complexification according to both of these conjectures has UV divergences, whereas the instantaneous energy is UV finite. This implies that neither the \textit{complexity equals action} nor \textit{complexity equals volume} conjecture is consistent with the conjectured bound on the rate of complexification.}

	\maketitle
	
	
	\section{Introduction} \label{intro}

	The complexity of a quantum state is defined as the minimum number of gates that map a reference state to {that quantum state} \cite{Susskind-comp,Susskind-comp-2,Aaronson:2016vto}. It was conjectured in \cite{Susskind-comp,Susskind-comp-2} that the growth of the complexity of a boundary CFT state as a function of time is holographically dual to the \textit{stretching} of the interior of a black hole in the bulk. 		
	To make this connection more concrete, various holographic definitions for the complexity have been put forward. One such proposal, known as the \textit{complexity equals action} (CA) conjecture, relates the holographic complexity with the on-shell gravitational action of a certain bulk region \cite{ca-1,ca-2}. This region, called the Wheeler-DeWitt (WDW) patch, is defined as the domain of dependence of a bulk Cauchy surface.  More precisely, this conjecture states that the complexity of a CFT state at time $\T$ is related to the on-shell action of the WDW patch corresponding to time $\T$, $\A(\T)$, as \cite{ca-1,ca-2}
	\begin{equation}
		\C_{A}(\T) = \frac{\A(\T)}{\pi} \, . \label{eq-CA}
	\end{equation}
	Another such proposal relates the complexity with the volume of an extremal Cauchy surface \cite{Stanford:2014jda,Susskind:2014jwa}. In particular, it is conjectured that the complexity of a CFT state at time $\T$ is related to the volume of an extremal Cauchy surface anchored on the boundary at time $\T$, $V_{\text{ext}}(\T)$, according to \cite{Stanford:2014jda}
	\begin{equation}
	\C_{V}(\T) \equiv \, \frac{V_{\text{ext}}(\T)}{G \ell_\text{AdS}} \, . \label{eq-CV}
	\end{equation}
	This proposal is called the \textit{complexity equals volume} (CV) conjecture. 
	
	
	The earliest checks \cite{ca-1,ca-2} for these conjectures were the observations that the rate of growth of these holographic complexities at late times are consistent with a universal bound by Lloyd. The Lloyd bound, which follows from the Margolus-Levitin theorem \cite{ML}, conjectures that the rate of complexification is bounded from above by the energy of the system \cite{lloyd}. That is,
	\begin{equation}
	\frac{d}{d\T} \C(\T) \le \, \frac{2}{\pi}  E \, , \label{eq-Lloyd}
	\end{equation}
	where $E$ is the instantaneous average energy of the system. The consistency of the holographic conjectures for complexity, Eqs.~\eqref{eq-CA}-\eqref{eq-CV}, with the Lloyd conjecture, Eq.~\eqref{eq-Lloyd}, has been verified in various contexts, \emph{e.g.}, see \cite{Yang:2016awy,Qaemmaqami:2017lzs,moosa,HosseiniMansoori:2017tsm}. However, some cases have been identified where the CA conjecture and the Lloyd bound are not compatible with each other \cite{Carmi:2017jqz,Kim:2017qrq,Couch:2017yil}. Moreover, it has been argued that the holographic gates for which the CA conjecture is true do not satisfy the assumptions of the Lloyd bound \cite{Cottrell:2017ayj}. In particular, these holographic gates are infinitesimal unitary transformations whereas the Lloyd bound is applicable for quantum gates that map a state into an orthogonal state \cite{Cottrell:2017ayj}. Therefore, there is no reason to expect that the holographic complexity conjectures satisfy the Lloyd bound. 
	
	In this paper, our goal is to present another example where the growth of the complexity computed using either the CA or the CV conjecture does not respect the Lloyd bound. More precisely, we consider starting from the vacuum state 
	of a $(3+1)$-dimensional CFT and perturb the Hamiltonian at $\T=0$ by a relevant operator with a time-dependent coupling. The Hamiltonian is given by
	\begin{equation}
		H(\T) = \, \HO + \lambda(\T) \, \OO \, , \label{eq-quench}
	\end{equation}
	where $\lambda(\T \le 0) = \, 0$. For simplicity, we only consider the case where the conformal dimension of the relevant operator is $\Delta = \, 3$. The subsequent state  evolves according to the perturbed Hamiltonian. 
	We claim that the complexity of the evolved state has time-dependent UV divergences. As a result, the rate of complexification is UV divergent. The instantaneous energy, on the other hand, is UV finite. Therefore, it does not even make sense to compare the rate of complexification with the instantaneous energy of the state as required by the Lloyd bound. For this reason, we consider this example to be a more violent violation of the Lloyd bound than those discussed in \cite{Carmi:2017jqz,Kim:2017qrq,Couch:2017yil}. 
	
	This time-dependent perturbation of a CFT is described using the AdS-CFT correspondence by the introduction of a (tachyonic) scalar field in the bulk \cite{Witten:1998qj,Klebanov:1999tb,Aharony:1999ti}. The conformal dimension $\Delta$ of the relevant operator fixes the mass whereas the coupling `constant' $\lambda(\T)$ fixes the boundary condition for this scalar field. The scalar field back-reacts on the bulk spacetime and changes the bulk geometry. Since the boundary condition for the scalar field is time-dependent, the bulk geometry is also time-dependent. Consequently, the on-shell action of the WDW patch and the volume of a Cauchy slice evolve as a function of time.
	
	The bulk geometry is determined by solving the Einstein-Klein-Gordon equations in the bulk. Finding the full solution of these equations is not an easy task. Luckily, the UV divergences in the on-shell action of the WDW patch and in the volume of a Cauchy slice only depend on the asymptotic geometry near the boundary \cite{comp-subregions,comp-div}. This means that it is sufficient to solve the bulk equations perturbatively in the radial coordinate. This method was also used in \cite{Hung:2011ta,Leichenauer:2016rxw} to investigate the UV divergences in the entanglement entropy. We perform this perturbative analysis in Sec.~(\ref{holo-setup}).
	
	We calculate the divergences in the rate of complexification in Sec.~(\ref{sec-CA}) and in Sec.~(\ref{sec-CV}). 
	In Sec.~(\ref{sec-CA}), we find the divergences in the time-derivative of the action of the WDW patch. This allows us to use the CA conjecture to find the divergences in the rate of complexification. In Sec.~(\ref{sec-CV}), we find the Cauchy slice with maximum volume anchored on the boundary at time $\T$, and extract the divergences from its volume. Combining this result with the CV conjecture yields the divergences in the rate of complexification. We find that the structure of the divergences in the complexity are the same as those identified in \cite{comp-subregions,comp-div} but the coefficients of these divergences depend on the instantaneous value of the coupling `constant', $\lambda(\T)$.
	
	The instantaneous energy of the boundary state is the expectation value of the Hamiltonian in Eq.~\eqref{eq-quench}. Like any other correlation function, the energy is a UV finite quantity once the renormalization procedure is performed. The response of the one-point function of the boundary stress-tensor to the time-dependent perturbation is studied using the AdS-CFT correspondence in \cite{Buchel:2012gw,Buchel:2013lla}. This requires variation of the on-shell renormalized action of the bulk theory with respect to the metric of the boundary. For our purposes in this paper, the details of the time-dependence of the energy is not significant. All we need to know for the violation of the Lloyd bound is the fact that the energy is a UV finite quantity.
	

	\section{Holographic Setup} \label{holo-setup}
	
	In this paper, we consider starting with a CFT (with a holographic dual) and deforming it by a relevant operator with a time-dependent coupling. The Hamiltonian of the perturbed theory is given in Eq.~\eqref{eq-quench}. The time-dependent perturbation of a CFT is studied extensively using the AdS-CFT correspondence in \cite{Leichenauer:2016rxw,Buchel:2012gw,Buchel:2013lla,Buchel:2013gba}. 
	According to the AdS-CFT correspondence, a relevant operator on the boundary is dual to a (tachyonic) scalar field in the bulk \cite{Witten:1998qj,Klebanov:1999tb,Aharony:1999ti}. This means that perturbing the boundary CFT by a relevant operator can be described by a scalar field in the bulk.
	The mass of this scalar field is fixed by the conformal dimension $\Delta$ of the relevant operator according to\footnote{We set $\ell_\text{AdS} = \, 1$ .} \cite{Witten:1998qj,Klebanov:1999tb,Aharony:1999ti}
	\begin{equation}
		m^{2} = \, \Delta\left(\Delta - 4\right) \, .
	\end{equation}
	For simplicity, we only focus on the relevant operator with conformal dimension $\Delta = 3$ in this paper. In this case, we have $m^2 = -3$.
	%
	
	This scalar field couples with the metric in the bulk, and the bulk theory is governed by the action \cite{Hung:2011ta,Leichenauer:2016rxw,Buchel:2012gw,Buchel:2013lla,Buchel:2013gba}
	\begin{align}
		\A\left[\Phi, g\right] =& \, \, \frac{1}{16\pi G} \int_{\M} d^{5}x \, \sqrt{-g} \Big( R + 12 - \frac{1}{2} \left(\partial\Phi\right)^{2} - V(\Phi) \Big) \nonumber\\  +& \, \, \frac{1}{8\pi G} \int_{\partial \M} d^{4}x \, \sqrt{-\gamma} \, K \, , \label{eq-action}
	\end{align}
	where 
	\begin{equation}
		V(\Phi) = \, -\frac{3}{2}\Phi^{2} + \sum_{n} \frac{1}{n!}  \kappa_{n}  \Phi^{n} \, .\label{eq-pot}
	\end{equation}
	The \textit{Newton's constant} that appears in the overall factor of the action in Eq.~\eqref{eq-action} can be fixed by demanding that the boundary two-point function of $\OO$ matches the holographic two-point function. That is, we can read off $G$ from \cite{Witten:1998qj,Klebanov:1999tb,Aharony:1999ti} 
	\begin{equation}
		\langle \OO(x)\OO(y) \rangle = \, \frac{1}{2\pi^3 G} \, \frac{1}{|x-y|^6} \, . \label{eq-def-G}
	\end{equation}  
	Similarly, the coupling constants $\kappa_{n}$ in Eq.~\eqref{eq-pot} can be determined by comparing the boundary $n$-point functions with the holographic $n$-point functions. 
	The term in the second line of Eq.~\eqref{eq-action} is the standard Gibbons-Hawking-York (GHY) boundary term without which the gravitational action does not have a well-defined variation \cite{Gibbons:1976ue}. No such boundary term is required for the scalar field with the Dirichlet boundary conditions \cite{Minces:1999eg}. 
	
	The on-shell value of the action in Eq.~\eqref{eq-action} has UV divergences, which arise because the volume element diverges near the asymptotic boundary \cite{Bianchi:2001kw}. To regulate these divergences, we need to introduce a cut-off surface near the asymptotic boundary. To make a choice of the cut-off surface, we choose to work with the Fefferman-Graham coordinates in the bulk \cite{Graham:1999pm,Fefferman:2007rka}. The most general metric of the asymptotically local AdS spacetime in these coordinates is\footnote{We follow the index convention that the Greek letters ($\mu,\nu,...$) denote boundary coordinates and Latin letters ($a,b,...$) correspond to bulk coordinates. } 
	\begin{equation}
	ds^{2} = \, \frac{1}{z^{2}} \, \Big( dz^{2} + G_{\mu\nu}(z,x) \, dx^{\mu}dx^{\nu} \Big) \, , \label{eq-FG-metric}
	\end{equation}
	where $z = 0$ is the asymptotic boundary. In these coordinate system, a convenient choice of a cut-off surface is $z = \, \delta$.

	
	The back-reaction of the scalar field on the bulk geometry is determined using the Einstein-Klein-Gordon equations in the bulk. The variation of the action in Eq.~\eqref{eq-action} with respect to the bulk metric yields
	\begin{equation}
		R_{ab} + 4 \, g_{ab} - \frac{1}{2} \partial_{a}\Phi \, \partial_{b}\Phi - \frac{1}{3} \, g_{ab} \, V(\Phi) \, = \, 0 \, , \label{eq-Einstein}
	\end{equation}
	whereas the variation of the action with respect to the bulk scalar yields
	\begin{equation}
		\frac{1}{\sqrt{-g}} \, \partial_{a} \, \Big( \sqrt{-g} \, g^{ab} \partial_{b} \Phi\Big) - \frac{\delta}{\delta \Phi} V(\Phi)  = \, 0 \, . \label{eq-scalar}
	\end{equation}
	These equations should be solved simultaneously to determine the bulk metric in Eq.~\eqref{eq-FG-metric}. However, Eq.~\eqref{eq-FG-metric} is the most general metric of the asymptotically local AdS spacetime. Since the perturbation of the boundary CFT is \textit{homogeneous} and \textit{isotropic}, we expect the bulk metric to have translation and rotational invariance in the transverse directions. Therefore, we make the following ansatz for the bulk metric\footnote{We denote bulk time coordinate by $t$ and boundary time coordinate by $\T$.} \cite{Leichenauer:2016rxw}
	\begin{equation}
	ds^2 = \, \frac{1}{z^{2}} \, \Big( dz^2 - f(z,t) \, dt^2 + \, h(z,t) \, \sum_{i=1}^{3} \, dx_{i}^{2} \Big) \, , \label{eq-metric}
	\end{equation}
	with the boundary conditions
	\begin{align}
	f(z=0,t) =&\, h(z=0,t) = \, 1 \, ,\\
	f(z,t\le 0) =&\, h(z,t\le 0) = \, 1 \, .
	\end{align}
	
	Finding the full bulk geometry and the profile of the bulk scalar field is a difficult exercise. However, as emphasized in Sec.~(\ref{intro}), we are only required to solve for the bulk metric and the bulk scalar near the asymptotic boundary. This allows us to expand the metric and the scalar field as a power series in $z$ and solve for the bulk equations perturbatively.	These perturbative solutions were studied in detail in \cite{Hung:2011ta}, where the powers of $z$ that appear in the asymptotic solutions of the bulk metric and the scalar field were identified. Furthermore, it was shown in \cite{Hung:2011ta} that the power series solution for the scalar field breaks down at order $z^{\Delta}$ and that for the bulk metric breaks down at order $z^{d}$, where $d$ is the dimensions of the boundary. Using these results, we write for $d=4$ and $\Delta = 3$
	\begin{align}
	f(z,t) =& \, \, 1 + z^{2} \, f_{2}(t) + z^{3} \, f_{3}(t) + z^{4} \, F(z,t) \, ,\label{eq-f}\\
	h(z,t) =& \, \, 1 + z^{2} \, h_{2}(t) + z^{3} \, h_{3}(t) + z^{4} \, H(z,t) \, ,\label{eq-h}\\	
	\Phi(z,t) =& \, \, z \, \phi_{0}(t) + z^{2} \, \phi_{1}(t) + z^{3} \, \Psi(z,t) \, ,\label{eq-phi}
	\end{align}
	with the boundary condition \cite{Witten:1998qj,Klebanov:1999tb,Aharony:1999ti}
	\begin{equation}
	\phi_{0}(t) = \, \lambda(t) \, . \label{eq-bc}
	\end{equation}
	
	As we will see in Sec.~(\ref{sec-CA}) and in Sec.~(\ref{sec-CV}), the non-perturbative part of the solution, that is $F(z,t)$ in Eq.~\eqref{eq-f}, $H(z,t)$ in Eq.~\eqref{eq-h}, and $\Psi(z,t)$ in Eq.~\eqref{eq-phi}, do not contribute to the divergences of either the action of the WDW patch or the volume of a Cauchy slice. So for our purpose, we ignore the non-perturbative part of the solutions and insert the perturbative ansatz in the bulk equations. Expanding the Klein-Gordon equation, Eq.~\eqref{eq-scalar}, to leading order in $z$ and using Eq.~\eqref{eq-bc} yield
	\begin{equation}
	\phi_{1}(t) =\, - \frac{\kappa_{3}}{2} \, \phi_{0}^{2}(t) \, = \, - \frac{\kappa_{3}}{2} \, \lambda^{2}(t) \, .\label{eq-phi1}
	\end{equation}
	Similarly, expanding the $zt$-component and $tt$-component of the Einstein's equations, Eq.~\eqref{eq-Einstein}, to the leading order yields
	\begin{equation}
	f_{2}(t) \, =\, h_{2}(t) \, = \, - \frac{1}{12} \, \lambda^{2} (t) \, , \label{eq-f2}
	\end{equation}
	whereas expanding these equations to next to the leading order yields
	\begin{equation}
	f_{3}(t) \, =\, h_{3}(t) \, = \, \frac{2}{27} \, \kappa_{3} \, \lambda^{3} (t) \, . \label{eq-f3}
	\end{equation}
	
	This finishes our discussion of the time-dependent bulk geometry due to the back-reaction of the scalar field. As we will see in the next two sections, the asymptotic results in Eqs.~\eqref{eq-phi1}-\eqref{eq-f3} are sufficient to calculate the divergences in the rate of complexification.

	\section{Complexity using CA conjecture} \label{sec-CA}
	
	In this section, we calculate the UV divergences in the rate of complexification using the CA conjecture. That is, we extract the time-dependent UV divergences that appear in the action of the WDW patch corresponding to the boundary time $\T$. 
	Although the details  of the WDW patch depend on the geometry far into the bulk, we are only concerned with the structure of the patch near the asymptotic boundary. We find the WDW patch near the asymptotic boundary in Sec.~(\ref{sec-wdw-patch}) and study its action in Sec.~(\ref{sec-wdw-action}).
	
	\subsection{Wheeler-DeWitt patch near the asymptotic boundary} \label{sec-wdw-patch}
	
	The WDW patch corresponding to the boundary time $\T$ is defined as the domain of dependence of a bulk Cauchy slice anchored to the boundary at time $\T$. This means that near the asymptotic boundary, the WDW patch is bounded by two null hypersurfaces, $\dot{J}^{\pm}(\T)$. The null hypersurface $\dot{J}^{+}(\T)$ ($\dot{J}^{-}(\T)$) is the boundary of the future (past) of the time slice $\T$ on the asymptotic boundary. Due to the translation invariance in the transverse directions, these codimension-$1$ null hypersurfaces can simply be described by
	\begin{equation}
	\dot{J}^{\pm}(\T) : \,\,\,\,\,\,\,\,\, t = t_{\pm}(z ;\T) \, , 
	\end{equation}
	with the boundary condition $t_{\pm}(z=0;\T) = \, \T$. Using the form of the bulk metric in Eq.~\eqref{eq-metric}, we deduce that the functions $t_{\pm}(z;\T)$ satisfy
	\begin{equation}
	\frac{d}{dz} t_{\pm}(z;\T) = \, \pm \frac{1}{\sqrt{f\big(z,t_{\pm}(z;\T)\big)}} \, .
	\end{equation}
	The perturbative solutions of these differential equations can be determined using the series expansion of $f(z,t)$ from Eq.~\eqref{eq-f}. Solving these equations order-by-order in $z$ and using Eqs.~\eqref{eq-f2}-\eqref{eq-f3} yields 
	\begin{equation}
	t_{\pm}(z;\T) = \, \T \pm z \, \pm \frac{1}{72} \, z^{3} \lambda^{2}(\T) \, + \frac{1}{96} \, z^{4} \left( \mp \frac{24 \, \kappa_{3}}{27} \lambda^{3}(\T) + \frac{d}{d\T}\lambda^{2}(\T) \right) \, + O(z^5) \, . \label{eq-tpm}
	\end{equation}
	
	As we discussed in Sec.~(\ref{holo-setup}), we introduce a cut-off surface near the asymptotic boundary at $z=\delta$ to regulate the UV divergences. This means that the WDW patch is also bounded by this cut-off surface. So to summarize, the WDW near the asymptotic boundary is the region bounded by the following inequalities
	\begin{equation}
	\text{WDW patch:} \,\,\,\,\,\,\,\,\,\,\,\,\,\, z \ge \, \delta \, , \,\,\,\,\, \text{and} \,\,\,\,\,\,\,\,\,\, t_{-}(z;\T) \le t \le t_{+}(z,\T) \, . \label{eq-wdw-region}
	\end{equation}
	
	After finding the WDW patch near asymptotic boundary, we study the action of this patch in the next subsection.	
	
	\subsection{Calculation of the Action} \label{sec-wdw-action}
	
	In addition to the action of the bulk theory given in Eq.~\eqref{eq-action}, the action of the WDW patch also has contributions from the null boundaries, $\dot{J}^{\pm}(\T)$, and from the corners where these null boundaries intersect the cut-off boundary, $z=\delta$ \cite{Lehner:2016vdi}. That is, the action of the WDW patch is given by the sum
	\begin{equation}
	\A(\T) = \, \A_{\text{bulk}}(\T) + \A_{\text{boundaries}}(\T) + \A_{\text{corners}}(\T) \, , \label{eq-wdw-action}
	\end{equation}
	where $\A_{\text{boundaries}}$ and $\A_{\text{corners}}$ include all the boundaries and all the corners respectively. 
	In the following, we first consider all the terms in Eq.~\eqref{eq-wdw-action} separately and then add them together to get the total action.
	
	The bulk contribution to the action of the region in Eq.~\eqref{eq-wdw-region} is given by
	\begin{equation}
	\A_{\text{bulk}}(\T) = \, \frac{1}{16\pi G} \int d^{3}x \, \int_{\delta} dz \, \int_{t_{-}(z;\T)}^{t_{+}(z;\T)} dt \, \sqrt{-g} \, \Bigg( R + 12 - \frac{1}{2} \left(\partial\Phi\right)^{2} - V(\Phi) \Bigg) \, .
	\end{equation}
	Using the trace part of the Einstein's equations, Eq.~\eqref{eq-Einstein}, and using the form of the metric in Eq.~\eqref{eq-metric}, we write the above expression as
	\begin{equation}
	\A_{\text{bulk}}(\T) = \, \frac{L^{3}}{16\pi G}  \, \int_{\delta} dz \, \frac{1}{z^5} \, \int_{t_{-}(z;\T)}^{t_{+}(z;\T)} dt \, f^{1/2}(z,t) \, h^{3/2}(z,t) \, \Bigg(-8 + \frac{2}{3} V(\Phi) \Bigg) \, , \label{eq-bulk-action-int}
	\end{equation}
	where we have also defined 
	\begin{equation}
	L^3 \equiv \int d^{3}x \, ,\label{eq-volume}
	\end{equation}
	as the volume of our boundary system. Evaluating the integrals in Eq.~\eqref{eq-bulk-action-int} is not possible without knowing an explicit form of the time-dependent coupling, $\lambda(\T)$. Nevertheless, we can still calculate the contribution of the bulk action to the rate of complexification. To do this, we take the time-derivative of bulk contribution in Eq.~\eqref{eq-bulk-action-int}. This yields
	\begin{align}
	\frac{d}{d\T} \A_{\text{bulk}}(\T) = \, \frac{L^{3}}{16\pi G}&  \, \int_{\delta} dz \, \frac{1}{z^{5}} \, \Bigg\{ \frac{d t_{+}(z;\T)}{d\T} \Bigg[f^{1/2}(z,t) h^{3/2}(z,t) \, \Big(-8 + \frac{2}{3} V(\Phi) \Big)\Bigg]\Bigg|_{t=t_{+}(z;\T)} \nonumber\\ -& \frac{d t_{-}(z;\T)}{d\T} \Bigg[f^{1/2}(z,t) h^{3/2}(z,t) \, \Big(-8 + \frac{2}{3} V(\Phi) \Big)\Bigg]\Bigg|_{t=t_{-}(z;\T)}  \Bigg\} \, . \label{eq-bulk-action-dot}
	\end{align}
	The integrand in the above expansion can be expanded as a power series in $z$ using Eqs.~\eqref{eq-f}-\eqref{eq-phi} and Eq.~\eqref{eq-tpm}. This allows us to extract the UV divergences in Eq.~\eqref{eq-bulk-action-dot}. Performing this analysis 
	and using Eqs.~\eqref{eq-bc}-\eqref{eq-f3} gives us the following time-dependent divergence in the bulk action
	\begin{align}
	\frac{d}{d\T} \A_{\text{bulk}}(\T) =& \, \frac{L^3}{36\pi G}\,\frac{1}{\delta} \, \frac{d}{d\T} \lambda^{2}(\T) \, . \label{eq-act-bulk-fin}
	\end{align}
	
	We now consider the boundary terms in the action, Eq.~\eqref{eq-wdw-action}. The region in Eq.~\eqref{eq-wdw-region} is bounded by three boundaries. One of these in the timelike cut-off surface at $z=\delta$, whereas the other two boundaries are null hypersurfaces, $\dot{J}^{\pm}(\T)$. We first consider the contributions from the null boundaries. The action of a null boundary is the integral of the \textit{surface gravity} of the null generator along the hypersurface \cite{Lehner:2016vdi}. This implies that the action of a null boundary vanishes if the corresponding null generator is affine-parameterized. The affine-parameterized null generators of the null boundaries, $\dot{J}^{\pm}(\T)$, are given by 
	\begin{equation}
	k_{\pm}^{a}(\T) = \, N_{\pm}(z;\T) \left(\pm z^{2} \big(\partial_{z}\big)^{a} + \frac{z^{2}}{f^{1/2}\Big(z,t_{\pm}(z;\T)\Big)} \, \big(\partial_{t}\big)^{a} \right) \, , \label{eq-null-vecs}
	\end{equation}
	where $N_{\pm}(z;\T)$ near the asymptotic boundary is given by
	\begin{equation}
	N_{\pm}(z;\T) = \, 1 + \frac{1}{24} z^{2} \lambda^{2}(\T) + O(z^3) \, .
	\end{equation}
	With this choice of the parameterization of the null generators, the null boundaries do not contribute to the action of the WDW patch. 
	
	We now focus on the timelike cut-off boundary. The action of this boundary is given by the standard GHY term
	\begin{equation}
	\A_{z=\delta}(\T) = \, \frac{1}{8\pi G} \, \int d^{3}x \, \int_{t_{-}(\delta;\T)}^{t_{+}(\delta;\T)} dt \, \sqrt{-\gamma} \, \gamma^{ab}\nabla_{a}s_{b} \, , \label{eq-action-cutoff}
	\end{equation}
	where 
	\begin{equation}
	s^{a} =\, -\delta (\partial_z)^a \, , \label{eq-normal}
	\end{equation}
	is the normal vector to the cut-off surface and $\gamma^{ab} \equiv g^{ab} -s^{a}s^{b}$ is the inverse induced metric on the boundary. Evaluating the integral in Eq.~\eqref{eq-action-cutoff} is not possible. Despite that, we can still calculate its contribution to the rate of complexification.  Taking the time-derivative of Eq.~\eqref{eq-action-cutoff} yields
	\begin{align}
	\frac{d}{d\T} \A_{z=\delta}(\T) = \, \frac{L^3}{8\pi G} \, \Bigg\{& \frac{d t_{+}(\delta;\T)}{d\T} \Big(\sqrt{-\gamma} \, \gamma^{ab}\nabla_{a}s_{b}\Big)\Big|_{t=t_{+}(\delta;\T)} \nonumber\\ & \,- \frac{d t_{-}(\delta;\T)}{d\T} \Big(\sqrt{-\gamma} \, \gamma^{ab}\nabla_{a}s_{b}\Big)\Big|_{t=t_{-}(\delta;\T)} \Bigg\} \, ,\label{eq-action-cutoff-der}
	\end{align}
	where we have also used Eq.~\eqref{eq-volume}. To extract the UV divergences in Eq.~\eqref{eq-action-cutoff-der}, we simply expand the right hand side in the power series of $\delta$. We get
	\begin{equation}
	\frac{d}{d\T} \A_{z=\delta}(\T) = \, -\frac{5L^3}{72\pi G} \, \frac{1}{\delta} \, \frac{d}{d\T} \lambda^{2}(\T) \, . \label{eq-act-bdy-fin}
	\end{equation}
	
	Lastly, we consider the corner terms in the action, Eq.~\eqref{eq-wdw-action}. The corners that contribute to the divergences in the action of the WDW patch are the codimension-$2$ surfaces where the cut-off surface intersect the null boundaries, $\dot{J}^{\pm}(\T)$. The action of these two corners is \cite{Lehner:2016vdi}
	\begin{equation}
	\A_{\text{corners}}(\T) = \, - \frac{1}{8\pi G} \, \int d^{3}x \, \sqrt{q_{+}} \, a_{+}(\T) - \frac{1}{8\pi G} \, \int d^{3}x \, \sqrt{q_{-}} \, a_{-}(\T) \, , \label{eq-action-cor}
	\end{equation}
	where the volume element is given by
	\begin{equation}
	\sqrt{q_{\pm}} = \, \frac{1}{\delta^3} \, h^{3/2}\Big(\delta,t_{\pm}(\delta;\T)\Big) \, = \frac{1}{\delta^3} - \frac{1}{8} \, \lambda^{2}(\T) \,  \frac{1}{\delta}  \, + O(1) \, , \label{eq-ind-metr-corner}
	\end{equation}
	and \cite{Lehner:2016vdi}
	\begin{equation}
	a_{\pm}(\T) = \, \log|s\cdot k_{\pm}| \Big|_{z=\delta \, ,\,  t=t_{\pm}(\delta;\T)} \, .
	\end{equation}
	Using the null vectors from Eq.~\eqref{eq-null-vecs} and normal vector from Eq.~\eqref{eq-normal}, we get
	\begin{equation}
	a_{\pm}(\T) = \, \log\delta + \frac{1}{24} \, \lambda^{2}(\T) \,  \delta^2  \, + O(\delta^3) \, .
	\end{equation}
	We combine this result with Eq.~\eqref{eq-ind-metr-corner} to simplify Eq.~\eqref{eq-action-cor} as\footnote{The divergences of the form $\frac{\log\delta}{\delta^3}$ and $\frac{\log\delta}{\delta}$ can be removed by adding a counter term to the action of the null boundaries \cite{comp-div}. This counter terms was originally introduced in \cite{Lehner:2016vdi} to ensure that the total action is independent of the overall normalization of the null vectors.}
	\begin{equation}
	\A_{\text{corners}}(\T) = \, - \frac{L^3}{4\pi G} \, \Bigg( \frac{\log\delta}{\delta^3} -  \frac{1}{8} \, \lambda^{2}(\T) \,  \frac{\log\delta}{\delta} + \frac{1}{24} \, \lambda^{2}(\T) \,  \frac{1}{\delta} \Bigg) \, .
	\end{equation}
	We take the time-derivative of this result to get
	\begin{equation}
	\frac{d}{d\T} \A_{\text{corners}}(\T) = \, \frac{L^3}{32\pi G} \, \frac{\log\delta}{\delta} \, \frac{d}{d\T} \lambda^{2}(\T) - \frac{L^3}{96\pi G} \, \frac{1}{\delta} \, \frac{d}{d\T} \lambda^{2}(\T) \, . \label{eq-act-cor-fin}
	\end{equation}
	
	To get the time-dependent divergences in the total action, we add the contributions from the bulk in Eq.~\eqref{eq-act-bulk-fin}, cut-off boundary in Eq.~\eqref{eq-act-bdy-fin}, and corners in Eq.~\eqref{eq-act-cor-fin}. We get
	\begin{equation}
	\frac{d}{d\T} \A(\T) = \, \frac{L^3}{32\pi G} \, \frac{\log\delta}{\delta} \, \frac{d}{d\T} \lambda^{2}(\T) \, - \frac{5 L^3}{96\pi G} \, \frac{1}{\delta} \, \frac{d}{d\T} \lambda^{2}(\T) \, .
	\end{equation} 
	
	The CA conjecture, Eq.~\eqref{eq-CA}, then implies that the UV divergence in the rate of complexification at time $\T$ is
	\begin{equation}
	\frac{d}{d\T} \C_{A}(\T) = \, \frac{L^3}{32\pi^{2} G} \, \frac{\log\delta}{\delta} \, \frac{d}{d\T} \lambda^{2}(\T) \, - \frac{5 L^3}{96\pi^{2} G} \, \frac{\log\delta}{\delta} \, \frac{d}{d\T} \lambda^{2}(\T) \, . \label{eq-ca-der}
	\end{equation}
	This is one of the main results of this paper. This verifies our claim from Sec.~(\ref{intro}) that the rate of complexification of the state following the time-dependent perturbation of a CFT has UV divergences. As argued in Sec.~(\ref{intro}), this result violates the Lloyd bound, Eq.~\eqref{eq-Lloyd}. To see this, note that the instantaneous energy of the state is UV finite \cite{Buchel:2012gw,Buchel:2013lla}. Hence, it is meaningless to expect a bound between the UV divergent rate of complexification and the UV finite energy of the state.
	
	We end this section with an interesting observation. We note that the divergences in Eq.~\eqref{eq-ca-der} are independent of the coupling constants, $\kappa_{n}$, in the bulk scalar potential, Eq.~\eqref{eq-pot}. This means that the divergences in the rate of complexification only depend on the two-point function of the unperturbed CFT through Eq.~\eqref{eq-def-G}. In other words, the coefficients of the divergences in Eq.~\eqref{eq-ca-der} are determined solely by the central charge of our original CFT, and hence are \textit{universal}.
	
	\section{Complexity using CV conjecture} \label{sec-CV}
	
	In this section, we use the CV conjecture to calculate the time-dependent UV divergences in the complexity of the boundary state. 
	Our approach here is to first find an \emph{extremal} Cauchy surface anchored on the boundary at time $\T$ and then to extract the UV divergences in the volume of that surface. 
	
	Due to translation symmetry in the transverse directions, we consider a bulk Cauchy surface anchored on the boundary at time $\T$ that can be described by
	\begin{equation}
	\Sigma_{\T} : \,\,\,\,\,\,\,\,\, t = T(z ;\T) \, ,
	\end{equation}
	with the boundary conditions $T(z=0;\T) = \T$. The future-directed normal vector to $\Sigma_{\T}$ is
	\begin{equation}
	n^{a} = \, \mathcal{N} \, \Bigg(  (\partial_{t})^{a} + \, f\Big(z,T(z;\T)\Big) \, T'(z;\T) \, (\partial_{z})^{a} \Bigg) \, ,
	\end{equation}
	where prime denotes derivative with respect to $z$, and the normalization factor 
	\begin{equation}
	\mathcal{N}^{-1} = \, \frac{1}{z} \, f^{1/2}\Big(z,T(z;\T)\Big) \, \sqrt{1- \, f\Big(z,T(z;\T)\Big) \, \Big(T'(z;\T)\Big)^{2} \,} \, .
	\end{equation}
	The extrinsic curvature of $\Sigma_{\T}$ is defined as
	\begin{equation}
	K_{\Sigma_{\T}} \equiv \, h^{ab} \nabla_{a}n_{b} \, , \label{eq-ext-curv}
	\end{equation}
	where $h^{ab} \equiv g^{ab} + n^{a}n^{b}$ is the inverse induced metric on $\Sigma_{\T}$. Finding an extremal bulk Cauchy surface is equivalent to demanding that the extrinsic curvature in Eq.~\eqref{eq-ext-curv} vanishes. This yields the following equation
	\begin{align}
	0 =& \, \Bigg\{T''  + \frac{1}{2}\frac{f'}{f}T' + \frac{1}{2}\frac{\dot{f}}{f}T'^{2} + \, \left(1-f T'^{2}\right)\left( \frac{1}{2}\frac{f'}{f}T' + \frac{3}{2}\frac{h'}{h}T' + \frac{3}{2}\frac{1}{f}\frac{\dot{h}}{h} - \frac{4}{z}T'\right)\Bigg\} \,\Bigg|_{t =\, T(z;\T)} \, ,\label{eq-stupid}
	\end{align}
	where dot denotes the derivative with respect to the bulk time coordinate, $t$. Solving this equation for $T(z;\T)$ is not feasible. However, we are only interested in the profile of the extremal Cauchy surface near the asymptotic boundary. This means we only need to solve Eq.~\eqref{eq-stupid} perturbatively. Demanding a series solution for $T(z;\T)$ and solving Eq.~\eqref{eq-stupid} order-by-order yield
	\begin{equation}
	T_{\text{ext}}(z;\T) = \, \T - \frac{1}{32} \, z^{4} \, \frac{d}{d\T} \lambda^{2}(\T) \, + O(z^5) \, . \label{eq-T-ext}
	\end{equation}
	
	After finding the extremal surface near the asymptotic boundary, we now find the divergences that appear in its volume. 
	The volume of this extremal surface is given by
	\begin{equation}
	V_{\text{ext}}(\T) = \, \int d^{3}x \, \int_{\delta} dz \, \frac{1}{z^4} \, h^{3/2}\Big(z,T_{\text{ext}}(z;\T)\Big) \, \sqrt{1-f\Big(z,T_{\text{ext}}(z;\T)\Big) \Big(T'_{\text{ext}}(z;\T)\Big)^{2} \, } \, . \label{eq-vol-big}
	\end{equation}
	Since the $z$ dependence in Eq.~\eqref{eq-T-ext} appears at $z^{4}$ order, it does not contribute to the divergences in Eq.~\eqref{eq-vol-big}. Therefore, we simply replace $T_{\text{ext}}(z;\T)$ with $\T$ in Eq.~\eqref{eq-vol-big} and get
	\begin{equation}
	V_{\text{ext}}(\T) = \, L^{3} \, \int_{\delta} dz \, \frac{1}{z^4} \, h^{3/2}(z,\T) \, , \label{eq-vol}
	\end{equation}
	where we have also used Eq.~\eqref{eq-volume}. Using the series expansion in Eq.~\eqref{eq-h} and using Eqs.~\eqref{eq-f2}-\eqref{eq-f3}, we find that the divergences in the volume of the extremal Cauchy surface are
	\begin{equation}
	V_{\text{ext}}(\T) = \, \frac{L^{3}}{3} \, \frac{1}{\delta^{3}} - \frac{L^{3}}{8} \, \frac{1}{\delta} \, \lambda^{2}(\T) - \frac{L^{3} \kappa_{3}}{9} \, \log\delta \, \lambda^{3}(\T) \, .
	\end{equation}
	
	The CV conjecture, Eq.~\eqref{eq-CV}, then implies that the divergences in the complexity at time $\T$ are
	\begin{equation}
	C_{V}(\T) = \, \frac{L^{3}}{3 G} \, \frac{1}{\delta^{3}} - \frac{L^{3}}{8 G} \, \frac{1}{\delta} \, \lambda^{2}(\T) - \frac{L^{3} \kappa_{3}}{9 G} \, \log\delta \, \lambda^{3}(\T) \, , \label{eq-cv-final}
	\end{equation}	
	whereas, the divergences in the rate of complexification are
	\begin{equation}
	\frac{d}{d\T} \C_{V}(\T) = \, - \frac{L^{3}}{8 G} \, \frac{1}{\delta} \, \frac{d}{d\T} \lambda^{2}(\T) - \frac{L^{3} \kappa_{3}}{9 G} \, \log\delta \,  \frac{d}{d\T} \lambda^{3}(\T) \, . \label{eq-cv-der-final}
	\end{equation}
	This is another main result of this paper. This result shows that the the CV conjecture, Eq.~\eqref{eq-CV}, and the Lloyd bound, Eq.~\eqref{eq-Lloyd}, are not consistent with each other. This follows from the fact that the energy of the state after the time-dependent perturbation is UV finite \cite{Buchel:2012gw,Buchel:2013lla}. This means that one of the sides of the Lloyd bound has UV divergences whereas the other side does not. Hence, the bound is not satisfied.
	
	Note that, unlike the result from CA conjecture, the divergences in Eq.~\eqref{eq-cv-final} and Eq.~\eqref{eq-cv-der-final} also depend on $\kappa_{3}$. This means that the divergences from CV conjecture also depend on the three-point functions of our original CFT, and hence are not universal.
	
	\section{Discussion}
	
	Our goal in this paper was to present a simple example where the holographic complexity computed either using the CA conjecture, Eq.~\eqref{eq-CA}, or the CV conjecture, Eq.~\eqref{eq-CV}, has time-dependent UV divergences. The example that we studied was a time-dependent deformation of a CFT Hamiltonian by a relevant operator with the assumption that the system was initially in the ground state of the unperturbed CFT. The bulk description of this perturbation involves a scalar field which back-reacts on the bulk geometry \cite{Witten:1998qj,Klebanov:1999tb,Aharony:1999ti}. Due to this back-reaction, the bulk geometry becomes time-dependent. Since we were only interested in the divergences in the holographic complexity, we only had to calculate the back-reaction of the scalar field near the asymptotic boundary in Sec.~(\ref{holo-setup}). With the time-dependent geometry near the asymptotic boundary, we were able to use the CA conjecture in Sec.~(\ref{sec-CA}) to determine the time-dependent divergences in the rate of complexification. This result is given in Eq.~\eqref{eq-ca-der}. Similarly, we used the CV conjecture in Sec.~(\ref{sec-CV}) to find the divergences that appear in the complexity. This result is given in Eq.~\eqref{eq-cv-der-final}. We find that the structure of the divergences in Eq.~\eqref{eq-ca-der} and in Eq.~\eqref{eq-cv-der-final} is the same as the one identified in \cite{comp-subregions,comp-div}.
	
	The significance of our result is that it shows that neither the CA conjecture nor the CV conjecture are consistent with the Lloyd bound, Eq.~\eqref{eq-Lloyd}. The Lloyd bound demands that the average energy of the system bounds the rate of complexification from above. However, the energy of the system following the time dependent perturbation is a UV finite quantity. The time-dependence of the energy of the perturbed system has been studied using the AdS-CFT correspondence in \cite{Buchel:2012gw,Buchel:2013lla}. Since the rate of complexification according to both the CA conjecture and the CV conjecture is UV divergent, whereas the energy of the system is not, the Lloyd bound cannot be satisfied. This is the main message of this paper. Though, note that the inconsistency of the CA conjecture and the Lloyd bound were also recently discovered in \cite{Carmi:2017jqz,Kim:2017qrq,Couch:2017yil}.

	

	\vskip .3cm
	\indent {\bf Acknowledgments} 
	It is a pleasure to thank Ning Bao, Raphael Bousso, Adam Brown, and Pratik Rath for helpful discussions, and Chris Akers, Ning Bao, and Saad Shaukat for useful feedback on a draft of this manuscript. This work was supported in part by the Berkeley Center for Theoretical Physics, by the National Science Foundation (award numbers 1521446 and 1316783), by FQXi, and by the US Department of Energy under Contract DE-AC02-05CH11231.
	
	\bibliographystyle{utcaps}
	\bibliography{all}
\end{document}